# Conformational Equilibria in Monomeric α-Synuclein at the Single Molecule Level

Massimo Sandal[1], Francesco Valle[1,2,*], Isabella Tessari[3], Stefano Mammi[4], Elisabetta Bergantino[3], Francesco Musiani[1], Marco Brucale[1], Luigi Bubacco[3], and Bruno Samorì[1,2,*]

[1]Department of Biochemistry "G. Moruzzi", University of Bologna, Via Irnerio 49, 40126 Bologna (Italy); [2]National Center on Nanostructures and BioSystems at Surfaces (S3) INFM-CNR (Modena, Italy); [3]Department of Biology University of Padova (Italy); [4]Department of Chemical Sciences, University of Padova (Italy).

* To whom correspondence should be addressed. E-mail: francesco.valle@unibo.it (FV), samori@alma.unibo.it (BS).

**Abbreviations**: AFM, atomic force microscopy; αSyn, α-synuclein; FTIR, Fourier transform infrared; NUP, natively unfolded protein; PBS, phosphate buffered saline; SDS, sodium dodecyl-sulfate; SMFS, single molecule force spectroscopy; WLC, worm-like chain.

**Running head**: Conformational Equilibria of α-Synuclein.

**AUTHOR SUMMARY:**

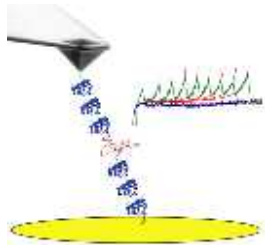

Natively unstructured proteins defy the classical "one sequence-one structure" paradigm of protein science. Monomers of these proteins in pathological conditions can aggregate in the cell, a process that underlies socially relevant neurodegenerative diseases such as Alzheimer and Parkinson. A full comprehension of the formation and structure of the so-called misfolded intermediates from which the aggregated states ensue is still lacking. We characterized the folding and the conformational diversity of αSyn, a natively unstructured protein involved in Parkinson disease, by mechanically stretching single molecules of this protein and recording their mechanical properties. These experiments permitted us to directly observe directly and quantify three main classes of conformations that, under *in vitro* physiological conditions, exist simultaneously in the αSyn sample, including disordered and "β-like" structures. We found that this class of "β-like" structures is directly related to αSyn aggregation. In fact, their relative abundance increases drastically in three different conditions known to promote the formation of αSyn fibrils: the presence of $Cu^{2+}$, the occurrence of the pathogenic A30P mutation, and high ionic strength. We expect that a critical concentration of αSyn with a "β-like" structure must be reached to trigger fibril formation. This critical concentration is therefore controlled by a chemical equilibrium. Novel pharmacological strategies can now be tailored to act upstream, before the aggregation process ensues, by targeting this equilibrium. To this end, Single Molecule Force Spectroscopy can be an effective tool to tailor and test new pharmacological agents.

**BLURB:**
A single molecule study detects structured and unstructured conformers in equilibrium in monomeric α-synuclein. The β-like conformers increase with pathological mutations and under other conditions known to promote aggregation.

**ABSTRACT**

Human α-Synuclein (αSyn) is a natively unfolded protein whose aggregation into amyloid fibrils is involved in Parkinson disease. A full comprehension of the structure and dynamics of early intermediates leading to the aggregated states is an unsolved problem of essential importance in deciphering the molecular mechanisms of αSyn aggregation and formation of fibrils.

Traditional bulk techniques utilized so far to solve this problem point to a direct correlation between the αSyn unique conformational properties and its propensity to aggregate but can only provide ensemble-averaged information for monomers and oligomers alike. They therefore cannot characterize the full complexity of the conformational equilibria that trigger the aggregation process.

We applied Atomic Force Microscopy-based single-molecule mechanical unfolding methodology to study the conformational equilibrium of human wild-type and mutant αSyn. The conformational heterogeneity of monomeric αSyn was characterized at the single molecule level. Three main classes of conformations, including disordered and "β-like" structures, were directly observed and quantified without any interference from oligomeric soluble forms. The relative abundance of the "β-like" structures significantly increased in different conditions promoting the aggregation of αSyn: the presence of $Cu^{2+}$, the pathogenic A30P mutation, and high ionic strength.

This methodology can explore the full conformational space of a protein at the single molecule level, detecting even poorly-populated conformers and measuring their distribution in a variety of biologically significant conditions. To the best of our knowledge, we present for the first time evidence of a conformational equilibrium that controls the population of a specific class of monomeric αSyn conformers, positively correlated with conditions known to promote the formation of aggregates. A new tool is thus made available to directly test the influence of mutations and pharmacological strategies on the conformational equilibrium of monomeric αSyn.

# INTRODUCTION

A significant fraction (possibly as much as 30%) of proteins and segments of proteins in eukaryotic proteomes has been found to lack, at least partially, a well-defined three-dimensional structure. Proteins belonging to this class are usually called natively unfolded proteins (NUPs) [1]. NUPs have been found to play key roles in a wide range of biological processes like transcriptional and translational regulation, signal transduction, protein phosphorylation and the folding of RNA and other proteins [2]. The conformational heterogeneity of NUPs allows them to adopt conformations that trigger pathogenic aggregation processes. In fact, NUPs are involved in the pathogenesis of some of the most widespread and socially relevant neurodegenerative diseases, such as Alzheimer and Parkinson [3-5]. Despite intensive research, the folding and the aggregation mechanisms of NUPs remain a major unsolved problem.

Theoretical studies depict the apparent structural disorder of NUPs as the result of the coexistence of a complex ensemble of conformers ensuing from a rugged energy landscape [6]. Five clusters of conformations, each with its own characteristic tertiary structure, were identified by Molecular Dynamics studies on the Alzheimer's β peptide [7]. Traditional bulk experiments and spectroscopies have recently been providing experimental evidence of the conformational diversity of these proteins [3,5]. Because of their inherent ensemble averaging, however, these methodologies cannot reveal the full complexity of the conformational equilibria of NUPs. Single molecule methodologies can single out the structures adopted by individual molecules within a complex conformational equilibrium [8-11,12,13,14].

We decided to approach the problem of the characterization of the conformers of α-synuclein (αSyn) (UniProtKB accession number P37840, SYUA_HUMAN), a prototype of this class of proteins. αSyn is a 140 aa protein expressed primarily at the presynaptic terminals in the central nervous system, and it is thought to be physiologically involved in endoplasmic reticulum - Golgi vesicle trafficking [15]. αSyn is involved in the pathogenesis of several neurodegenerative diseases, called synucleopathies. Intracellular proteinaceous aggregates (Lewy bodies and Lewy neurites) of αSyn are hallmarks of Parkinson disease [16] and multiple system atrophy [17]. Three naturally occurring mutations in the αSyn protein sequence, A30P, A53T, and E46K have been identified so far in human families affected by familial Parkinsonism [18-20]. These mutant proteins display an increased tendency to form non-fibrillar aggregates [21] and Lewy-bodies-like fibrils *in vitro* [22].

The fibrils spontaneously formed by αSyn by a nucleation-dependent mechanism are rich in β structure[23,24]. The transition from the natively unfolded monomeric state to fibril is therefore a process of acquiring structure. This process is still under strong debate. Evidence is accumulating that the monomeric αSyn, under *in vitro* physiological conditions, populates an ensemble of conformations including extended conformers and structures more compact than expected for a completely unfolded chain [25-32]. The marked differences between the scenarios depicted in those studies are mostly determined by the different time scales of the ensemble-averaging of the different methods utilized. Moreover, it is difficult for bulk methodologies to single out the monomeric state in the presence of soluble oligomers when they form quickly in solution [33]. On the contrary, the Single Molecule Force Spectroscopy (SMFS) approach reported here describes, by design, the conformational equilibrium of the monomeric form.

The different structures assumed by αSyn have been commonly investigated by adding to its buffer solution different chemicals, such as methanol or trifluoroethanol [34], metal cations like $Cu^{2+}$ and $Al^{3+}$ [35,36], or SDS micelles [37-39] in order to shift the conformational equilibrium towards the form under investigation. A previous force spectroscopy experiment showed that a relevant 12-amino acid segment of αSyn is conformationally heterogeneous [40]. The approach we report can span the full conformational space of the whole protein and also identify poorly populated conformers of the monomeric αSyn in *in vitro* physiological conditions. Three distinct classes of structures in equilibrium were identified: random coil, a mechanically weak fold, and "β-like". Their populations were also monitored under conditions known to influence aggregation, such as the presence of $Cu^{2+}$, high buffer concentration and, most importantly, the pathogenic mutation A30P.

**RESULTS**

In order to stretch an individual αSyn molecule by AFM we need handles to connect one end of the protein to the tip and the other to the substrate. To this end, we followed the design proposed by J. Fernandez for the study of the random coiled titin N2B segment [41]. A chimeric polyprotein composed of a single αSyn module flanked on either side by three tandem I27 domains (PDB entry: 1tit) (3S3, Fig. 1A) was expressed [42-44]. These domains act as molecular handles to mechanically stretch a single αSyn molecule. They also introduce well-characterized fingerprint signals into the recorded force curves that make it possible to identify the different αSyn conformations. The design is such that if the number of unfolding signals coming from I27 modules is larger than four, we are sure to have also mechanically stretched the αSyn module in the middle (see Fig. 1A). Among the curves showing mechanical unfolding events, however, only those featuring at least six unfolding peaks were selected and analyzed. This choice reduced the statistical sample even more, but it allowed us to recognize, in a very stringent way, the signatures of the different conformations of the αSyn moiety on each construct molecule that had been stretched.

To probe the native-like conformer population of αSyn, we performed experiments in a 10 mM Tris buffer solution. We found that the profiles of the selected force curves can be classified into three main classes. Two were unambiguously assigned to well-defined classes of conformers: one with the typical mechanical behavior of random-coil chains and the other of β-like structures. We propose that the profiles of the third class correspond to fairly compact architectures, likely to be sustained also by interactions among different modules of the construct.

**The signature of disordered conformers of αSyn with the mechanical behavior of an entropic random coil**

In the class of traces depicted in Fig. 1B, the force curve exhibits (from left to right) a long initial region, without any significant deviation from the Worm-Like Chain (WLC) behavior [45], followed by a saw-tooth pattern with six consecutive unfolding events, in addition to the last one that corresponds to the final detachment of the molecule from the tip. The initial region corresponds to the extension of a chain that occurs at low force and without significant energy barriers limiting its extensibility.

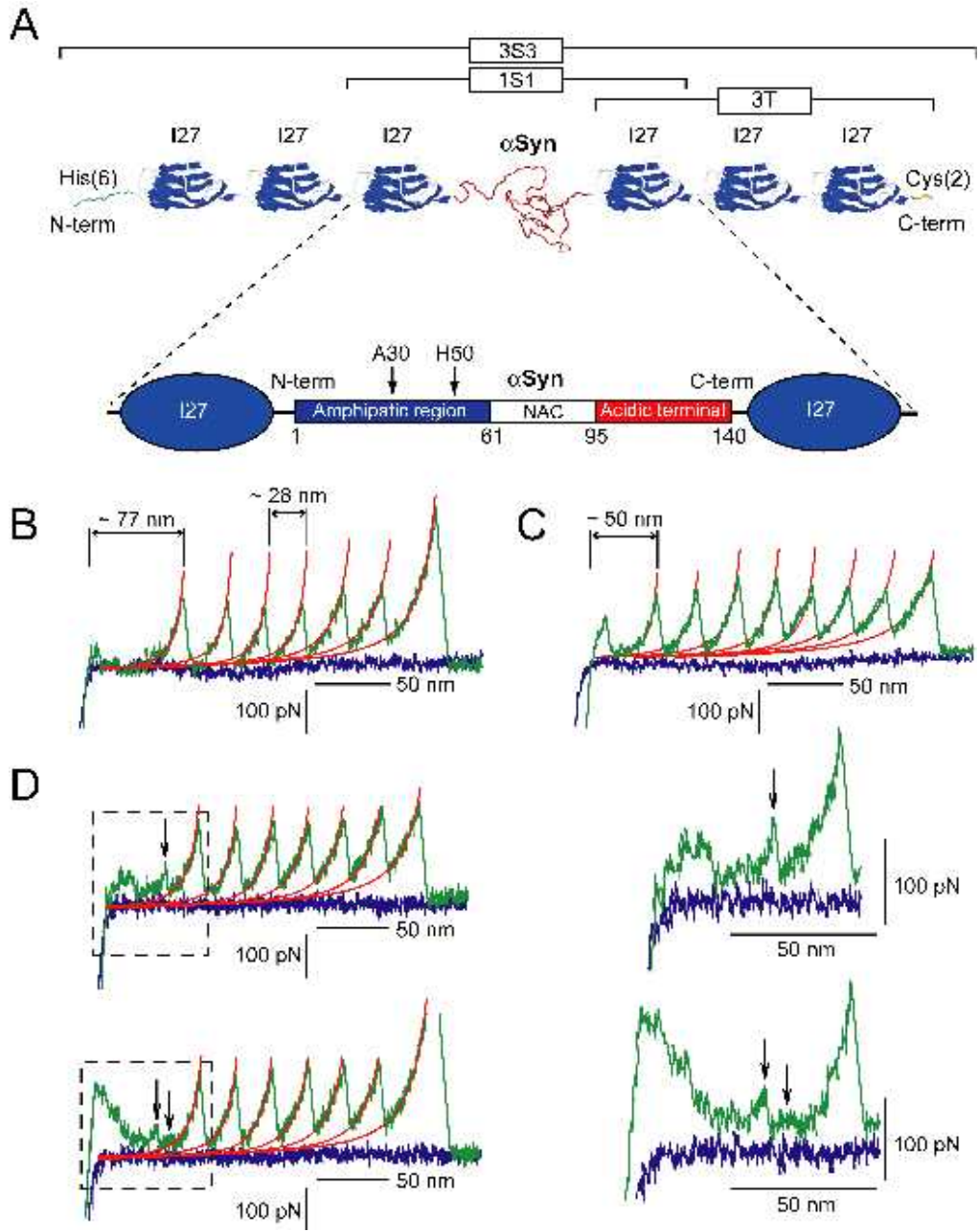

**Figure 1**. **The mechanical signatures of αSyn conformational classes as recorded by SMFS.** (**A**) Schematic representation of the polyprotein constructs used in this work: 3S3 contains the αSyn sequence (*red*) flanked on either side by three titin I27 modules (*blue*), the N-terminal His-tag, needed for purification purposes, (*green*) and the C-terminal Cys-Cys tail needed for covalent attachment to the gold surface (*yellow*); in 1S1, the αSyn moiety is flanked only by one I27 on both sides; the 3T is made up by three I27s. In the αSyn moiety (enlarged), three regions are shown: i) the amphipathic region, prone to fold in α-helical structures when in contact with phospholipid membranes; ii) the fibrillogenic NAC region, characteristic of the fibril core of αSyn amyloid; and iii) the acidic C-terminal tail, strongly charged and not prone to fold. The positions of alanine 30, site of the A30P mutation and histidine 50, crucial for the binding of $Cu^{2+}$, are marked. (**B**) Example of curve characterized by a featureless region assigned to the stretching of αSyn moiety having, in this case, the mechanical properties of a random coil (see text). This region is followed (from left to right) by six unfolding peaks of about 200 pN, with about 28 nm gaps between each other, assigned to the unfolding of I27 domains. (**C**) Example of the curves featuring the "β-like" signature of αSyn (see text), showing seven practically indistinguishable unfolding events of similar magnitude and spacing. (**D**) Curves featuring the signature of mechanically weak interactions, showing single or multiple small peaks (arrows) superimposed on the purely entropic WLC behavior of the trace preceding the six saw-tooth-like peaks. Right panels show a zoom of the region enclosed by the dashed squares.

The six unfolding peaks are spaced by ~28 nm. This spacing between the peaks corresponds to an 89 amino acid chain (0.36 nm per amino acid [46]), *i.e.*, to the increase in length of the protein after the unfolding of one I27 domain. These six unfolding peaks correspond to the characteristic fingerprint of the mechanical unfolding of the I27 modules [41]. We can therefore infer that, in this case, the AFM tip picked up the 3S3 construct molecules at the His-tag terminus, while the other end was tethered to the gold surface by the C-terminal cysteines.

The location of the first unfolding peak of I27, corresponding to the contour length of the construct molecules prior to any unfolding event, proves that the preceding featureless part of the trace can be unambiguously assigned to the αSyn chain. In fact, the measured contour length that fits this peak is 77 ± 4 nm. Subtracting the length of the six, still folded, I27 domains from this value (4.5 nm each [47]), a value of 48 ± 4 nm is obtained. This length corresponds to the chain of 140 aa of the αSyn. Therefore, this featureless initial part is the signature of αSyn conformers with the mechanical properties of a random coil. Their average persistence length was estimated by fitting the WLC model at 0.36 ± 0.05 nm. About 38% of the molecules showed this mechanical behavior in Tris/HCl buffer 10 mM (Fig. 2).

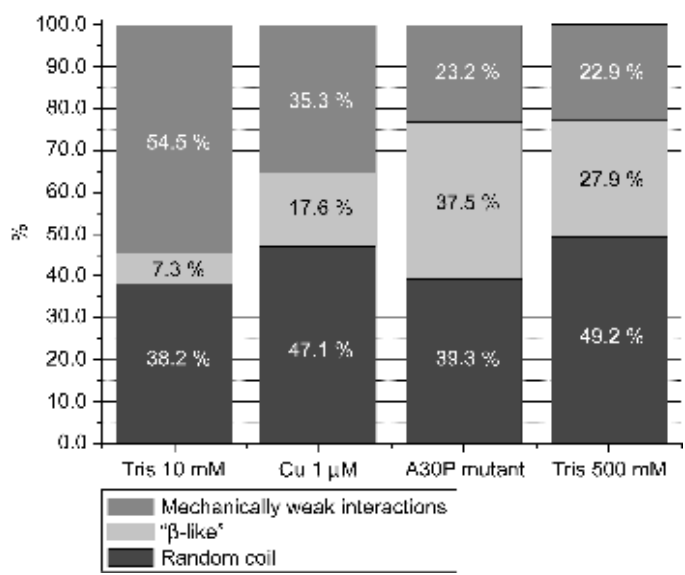

**Figure 2**. **Population shift of αSyn conformers in different conditions.** Population of αSyn conformers in the four different conditions tested in the present work. Percentages observed for each curve type (see Fig. 1) at 10 mM Tris/HCl (n = 55), 10 mM Tris/HCl with 1 μM $Cu^{2+}$ (n = 34), the A30P mutant in 10 mM Tris/HCl (n = 56) and 500 mM Tris/HCl (n = 61).

**The signature of αSyn conformers with the mechanical behavior of a chain containing a β-like structured segment**

A significant proportion of force curves with seven regularly spaced unfolding peaks in the 200 pN range (in addition to the last one corresponding to the final detachment) (Fig. 1C) was also recorded.

The presence of a number of unfolding events greater than that of the I27 modules in the construct cannot be ascribed to a possible simultaneous pulling of more than one 3S3 molecule because pulling two multi-domain constructs at the same time would not likely lead to a uniform separation between the I27 unfolding events. Moreover, we never obtained a significant and uniform set of reproducible curves with eight or more unfolding peaks with 28 nm separation.. Curves with seven unfolding events were well reproducible and their statistics was unambiguously modulated by conditions able to trigger aggregation, for example, ionic strength, the presence of $Cu^{2+}$ ions and, most importantly, pathogenic mutations (see below). The appearance of seven unfolding events cannot come from a construct accidentally expressed with seven, instead of six, I27 domains because of the cloning strategy (see Materials and Methods section). The occurrence of a seventh peak due to the stretching of 3S3 dimers can be also ruled out. Dimers could form in solution via disulfide bonds between the terminal cysteines, but those bonds tend to dissociate into thiols in the presence of gold, because the gold-sulfur bond is more stable than the sulfur-sulfur bond [48]. Each monomer contains two terminal cysteines and one of them could be involved in the dimerization and the other could bind to the gold surface. Even in this unlikely event, the length of the tethered chains extending from the surface is the same as that of a non-dimerized construct. Therefore, also in the case of a dimer tethered to the surface, more than six I27 unfolding peaks with the same separation cannot be recorded. We nevertheless tested the sample using DTT to avoid any disulfide-bonded dimer formation. Under these conditions, the statistics of different populations was comparable to those in the standard buffer, and we still recorded a significant proportion (~10%) of 7-peaked curves.

Because of the previous considerations, we therefore assign one of the seven peaks to the unfolding of the αSyn moiety. The length (95 aa) of this αSyn "β-like" folded section accidentally coincides with that of the I27 domain. This coincidence hinders the possibility to discriminate the peak of the αSyn from the six of the I27 domains. Nevertheless, the assignment of these curves to the unfolding of the αSyn moiety is confirmed by the position of the first unfolding peak, *i.e.,* by the contour length of the construct molecules prior to any unfolding event. As shown in Fig. 3, the position values correspond to a chain composed of the six I27 folded modules, plus the αSyn moiety with its C-terminal segment of 50 aa fully unfolded, and the remaining 95 amino acids folded into a structure with the same contour length as a folded I27 domain (solid line). The low propensity to fold of the 50 aa of the very acidic αSyn C-terminal tail has been extensively documented [38,39,49]. Segments of the remaining 95 amino acids are instead known to fold under different conditions into an α-helix [38,39] or, in the amyloid, into a β-sheet structure [31]. It must be noted that in about 40% of the molecules, the contour length of the same folded section is larger than that corresponding to 95 aa. The αSyn structural diversity therefore includes also "β-like" chain portions with different lengths. This interpretation is confirmed by comparing the variance of the folded section of 7-peaked curves with that of I27 modules (Fig. 3).

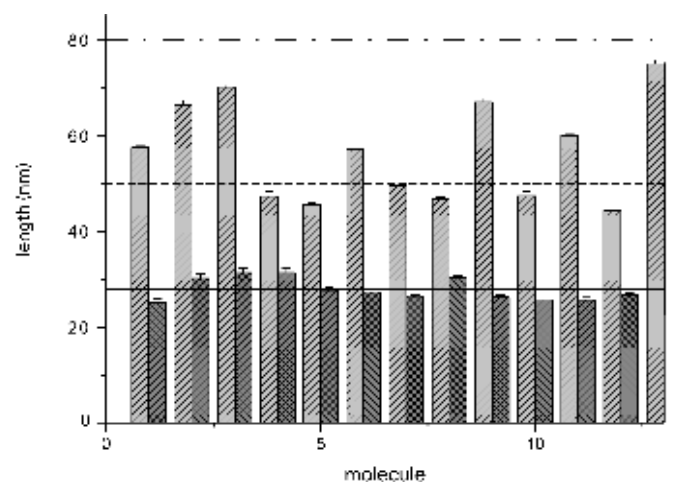

**Figure 3. Contour length analysis of "β-like" force curves.** Values of the first peak position in force curves showing seven unfolding peaks. The height of each bar corresponds to the initial contour length of a single curve, obtained by fitting the first unfolding peak by means of the WLC model. The dashed line is the length corresponding to a protein construct with six I27 folded modules plus 95 aa of αSyn folded into a β-like structure, and the remaining 50 aa of αSyn unstructured (see text). The dashed-and-dotted line is the length corresponding to a protein construct with six I27 modules plus the 140 aa of αSyn completely unstructured. The lengths of twelve randomly chosen I27 modules have also been reported (dark gray columns) for comparison. The solid line is the nominal I27 contour length. The larger spreading of the αSyn data confirms the higher conformational heterogeneity. Side quotas show the difference between the maximum and minum observed length value for I27 (bottom) and "β-like" structures (top).

The 200 pN unfolding force of all the seven peaks indicates that the folded section of αSyn has the same mechanical properties of the I27 β-sandwich structure. At the moment, without any independent structural characterization, we consider and label this folded structure of the αSyn moiety just as "β-like", in accordance with its mechanical behavior. Nevertheless, its mechanical behavior is in agreement with a β-sheet content in the "β-like" class of conformers. It is unlikely that the α-helical content we observed by means of CD (see below) correlates with the "β-like" conformers. In fact, whereas β-structures, like those of titin modules, such as I27 [41,50], or tenascin [51], unfold at forces in the range of 100-300 pN (at loading rates of the order of $10^{-5}$ N/s), the α-helix domains, in the same conditions, are always observed to unfold at forces almost one order of magnitude smaller [52-55].

In conclusion, these curve profiles provide clear evidence that in 10 mM Tris/HCl buffer, about 7% of the molecules (Fig. 2) contain a segment of the αSyn chain of about 95 aa folded into a structure with the mechanical property of the I27 β-sandwich structure. This percentage of the "β-like" structures, as we will see below, can be related with conditions leading to pathogenic aggregation.

**Signatures of further conformational complexity probed as mechanically weak interactions**

The remaining force spectroscopy curves (Fig. 1D) show single or multiple small peaks (sometimes with a plateau- or dome-like appearance) superimposed on the purely entropic WLC behavior of the trace preceding the six saw-tooth-like peaks.

The geometry of our construct made it possible to exclude that those small peaks might correspond to the rupture of aspecific αSyn-gold interactions. In fact, if the unstructured αSyn was adsorbed on the surface, upon pulling the construct, we would have recorded the first event at a distance from the tip contact point corresponding to the length of the three I27 modules (ca. 13.5 nm). The mechanically weak events we observed instead took place at an average distance from the contact point of 60 ± 26 nm with no events below 20 nm. They are therefore not compatible with αSyn-gold interactions. We assign these signals to the rupture of mechanically weak interactions placed at short- and long-distances along the chain.

The average forces of those single or multiple small peaks of the profiles are in the 64 ± 30 pN range (well above the noise level), without a defined hierarchy; often stronger peaks precede weaker ones, hinting topologically "nested" interactions. From the difference between the contour length estimated at those small peaks and that at the first I27 unfolding peak, one can measure the size of the topological loop enclosed by the interactions whose rupture is monitored by the different peaks. The resulting broad distribution of these distances monitors the ample multiplicity of these interactions as discussed in the Supplementary Material.

More than 50% of the molecules showed short- and long-distance mechanically weak interactions in 10 mM Tris/HCl buffer (Fig. 2). These interactions were also monitored by ensemble-averaged fluorescence spectroscopy. The fluorescence comes from the tryptophan residues of the I27 domains which are absent in αSyn. The fluorescence spectra reported in Fig. 4A prove that interactions between the I27 handles and tracts of the αSyn moiety do take place, as shown by the broadening of the spectrum of 1S1 with respect to that of the 3T construct (See Fig. 1A and Materials and Methods section for constructs description.) and by the 5 nm shift of the $\lambda_{max}$. The possibility of partial I27 unfolding leading to Trp exposure and broadening of the spectrum is ruled out by the CD data and by our force curves, which show that I27 domains are as tightly folded in the 3S3 construct as in an I27 homopolymer. A broadening due to subtle conformational effects on the I27 domain that expose the I27 Trp residue is possible, but even in this case, the fact that this broadening happens only when the αSyn moiety is inserted in the construct proves that direct interaction is taking place.

CD spectra of 1S1 and 3T were recorded in which 1S1 shows some α-helical content in the αSyn moiety (Fig. 4B). . Subtraction of the contribution of the I27 linkers (2/3 of the CD of 3T recorded in the same 10 mM Tris/HCl buffer) from the CD spectrum of 1S1 reveals a profile that is different from that of αSyn in the same buffer condition (Fig. 4C) but similar to that of the same protein in the α-helix structure induced by the addition of SDS [33]. This α-helical content might be induced by the interactions between the αSyn moiety and the I27 domains as discussed below and in the Supplementary Material.

**Low concentrations of $Cu^{2+}$ affect the conformational equilibrium of αSyn**

It is well known that multivalent metal cations like $Cu^{2+}$ can accelerate αSyn aggregation [35,36]. To validate our approach and to investigate how metal cations influence the conformer equilibrium of αSyn, we performed SMFS experiments on the 3S3 construct in 10 mM Tris/HCl buffer in the presence of 1 μM $CuCl_2$. The low concentration of copper was chosen to target the His 50 specific copper binding site of αSyn ($K_d$ = 0.1 μM) [36].

The presence of 1 μM $Cu^{2+}$ moderately, but significantly ($\chi$-square statistical significance $p$ < 0.01), alters the relative distribution of the αSyn conformers with respect to plain 10 mM Tris/HCl (see Fig. 2). In particular, the relative population of the β-like conformers more than doubles (from 7,2% to almost 18%), with a parallel decrease of the signals coming from mechanically weak interactions. An increase (from 38% to 47%) of random coil-like curves is also observed.

**The conformational equilibrium of the pathological A30P αSyn mutant is drastically shifted towards the β-like conformers**

The A30P mutation is a pathogenic, naturally occurring human αSyn variant, that correlates with familial Parkinsonism (OMIM reference number 163890.0002) [19]. The mutant protein displays an increased rate of oligomerization [56] and impaired degradation by chaperone-mediated autophagy [57]. We tested the 3S3 αSyn-A30P construct to evaluate the capability of our methodology to probe different conformational propensities in mutants of the same protein. We found that the A30P mutation induces a striking shift in the conformational equilibrium of αSyn with "β-like" curves being around 37% of the sample and again, a corresponding decrease of signals coming from mechanically weak interactions (Figure 2). In contrast with wild type αSyn incubated with $Cu^{2+}$, the A30P mutant does not induce an increase of random coil curves that are exactly in the same proportion observed in wild-type αSyn.

**The relative population of the three classes of conformers is modified by the buffer concentration**

Another condition known to speed up αSyn aggregation is high ionic strength[26][28]. SMFS experiments on the 3S3 wild type construct were performed in 500 mM Tris/HCl buffer. As reported in Fig. 2, the frequency of the three types of profiles radically changed in those two conditions. The most remarkable result is, again, the significant increase in the population of the "β-like" structures with buffer concentration (up to about 28%) and the parallel decrease of the percentage of the mechanically weak structures. Again, an increase of random coil curves is also observed, as occurs in the presence of $Cu^{2+}$, but unlike the case of the A30P mutant.

**DISCUSSION**

We have identified the signatures of three classes of conformers in monomeric αSyn at the single molecule level. One of these classes includes structures mechanically indistinguishable from a random coil; the other two classes include "β-like" structures and structures kept together by short- and long-distance mechanically weak interactions (Fig.1). We have also observed that their equilibrium shifts significantly depending on solution conditions or sequence variants related to pathological aggregation. The important result that emerges from these data is the direct correlation between conditions known to increase the αSyn aggregation propensity and the relative size of the "β-like" population (Fig.2).

**The "β-like" conformers are structured conformations directly related to the aggregation propensity**

We observed a marked increase of the population of "β-like" conformers under three very different conditions known to accelerate αSyn aggregation. This result links the population of those αSyn monomeric conformers to the process of αSyn aggregation. The first condition is the presence of a μM concentration of $Cu^{2+}$. Our results in this condition agree with the observation of a metal-induced partially folded intermediate by Uversky, Li and Fink[35]. Also Rasia *et al.* suggested a compact set of metal-induced conformations, noticing that the specific binding of $Cu^{2+}$ to the αSyn N-terminus requires the formation of a metal-binding interface (pivoted on His 50), that possibly involves residues widely separated in the primary amino acid sequence[36].

The second condition is the A30P mutation. NMR experiments have observed a much more flexible average conformation of the αSyn mutants A30P and A53T. The increased average flexibility of αSyn allows the protein to sample a larger conformational space. [58]. Interestingly, the mean hydrodynamic radius of αSyn is not affected by the A30P and A53T mutations [21,59] thus showing that the increased flexibility is compatible with the population of compact folded structures like those singled out by our experiments.

The third condition is a radical increase of the ionic strength. Our results in 500 mM Tris/HCl can be reconciled with the model proposed by Hoyer *et al.* [26] and by Bernado *et al.* [28] to explain the well-documented phenomenon of the increased αSyn fibril formation with increasing ionic strength. According to that model, the increased fibril formation is explained just on the basis on an increased freedom of the fibrillogenic NAC region caused by the release of its interaction with the negatively charged C-terminal tail. The increased ionic strength of the buffer leads to a more efficient charge shielding of the strongly acidic C-terminal tail, thus relieving its electrostatic self-repulsion. This in turn leads to the lowering of the protein-excluded volume and increases its flexibility. According to our data in Fig. 2, we should add to this model a shift of the conformational equilibrium towards the β-like structures that takes place on increasing the charge shielding.

**Are "β-like" structures really β?**

Any assignment of force spectroscopy signals to a definite secondary canonical structure must be supported by independent structural data. We have labeled as "β-like" those conformers with a mechanical behavior closely matching those of structures rich in β-sheets. The correlation of the population of these structures with aggregation conditions, that enrich β-sheet content in αSyn, supports this labeling. Evidence of some β-sheet content in the monomeric state of αSyn was previously reported in the literature. Most recently by means of NMR spectroscopy in supercooled water at minus 15 C°, it was found that the αSyn chain, cold-denatured to an hydrodynamic radius equivalent to that displayed by the same protein in 8 M urea, retains a surprising amount of unpacked β-strand content that correlates with the amyloid fibril β structure[32]. The packing of these β strands into compact structures like those observed by us is thus likely to occur in non-denaturing conditions and at physiological temperatures. This NMR result supports our observation of "β-like" conformers in the monomeric state of αSyn and links them to the amyloid β structure. The presence of β-sheet structures was indicated also by Raman spectra of this protein in aqueous solution[33]. In the same investigation, CD spectroscopy proved unable to detect any β content. Correspondingly, the CD spectra of αSyn recorded by us in 10 mM and 500 mM Tris were practically superimposable. We conclude that CD is not a technique sensitive enough to detect partial β-sheet content in the αSyn sample. A fraction of β-sheet/extended structure of about 19% was also detected, again not by CD, but by FTIR in dried films of αSyn[60]. This fraction is much larger than that estimated by our experiments in 10 mM Tris/HCl buffer (see Fig. 2). However, the conditions of the SMFS and FTIR experiments were markedly different, and in the latter case, some template-mediated formation of β-structures due to the packing of the αSyn molecules in the dried films required by the FTIR measurements cannot be ruled out.

In conclusion, despite the fact that force spectroscopy data cannot directly assign a specific secondary structure to the conformers we have labeled as "β-like", it is most likely that they have significant β-sheet content.

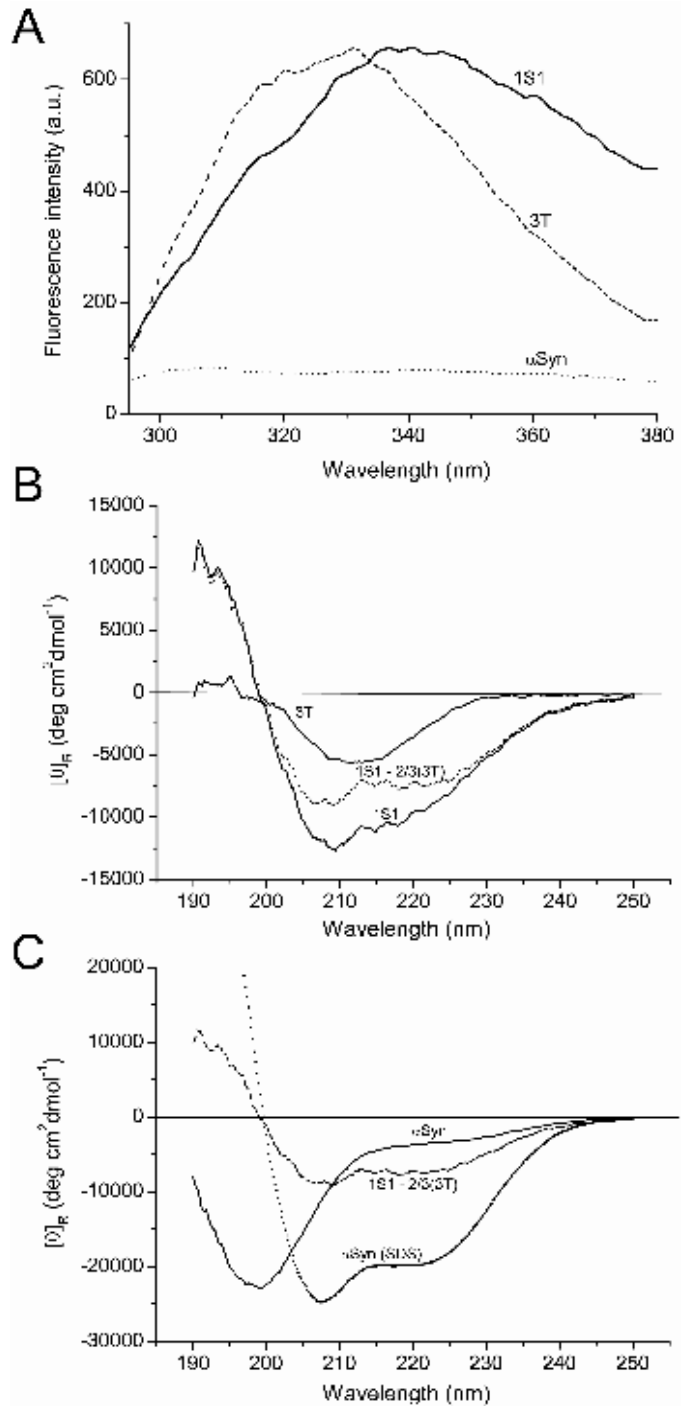

**Figure 4. Circular dichroism and fluorescence spectroscopy.** (**A**) Fluorescence spectra of αSyn, 3T and 1S1 (dotted, dashed and solid line, respectively). (**B**) Circular dichroism spectra in PBS buffer of 3T, 1S1 (solid lines). The αSyn contribution in 1S1 (dashed line) is calculated by subtracting the relative contribution of the I27 domains from the CD spectrum of 1S1. (**C**) Circular dichroism spectra of αSyn in PBS (solid line) and 250 mM SDS (dotted line). The αSyn contribution in 1S1 (dashed line) is reported as in **B**.

**The mechanically weak interactions can be both within the αSyn moiety and between α Syn and the I27 domains.**

By now, any structural characterization of the mechanically weak interactions events monitored by the small peaks in force curves as in Fig.1D (right panels) is at best tentative, and falls outside the focus of the present work. A more detailed characterization of these events is however within the range of capabilities of the techniques proposed here and is being currently addressed in our laboratory (see Supplementary Material for preliminary measurements). A plausible explanation of the short- and long-distance mechanically weak interactions we observed cannot exclude the interaction between positively charged residues on the αSyn N-terminal and the negatively charged surface of I27 modules[61]. It has been documented that αSyn in contact with negatively charged surfaces assumes an α-helix structure [37-39,62,63]. We might expect a similar structural transition in the αSyn moiety also from the contact with the I27 modules within the 3S3 or 1S1 constructs (see Supplementary Material). This transition is indicated by the CD spectra of the 1S1 construct in 10 mM Tris/HCl (see Figure 4B). We propose that the small peaks like those shown in Fig.1D and assigned to the mechanically weak interactions can be the signature of the interaction between αSyn, possibly in α-helical form, and the flanking I27 modules. It is not surprising that more than one of those signals are present in the same force curves, because multiple interactions of this type can occur at the same time in the same molecule. It should be noted that the same transition does not take place when free αSyn is mixed in solution with I27 modules of the 3T construct (see Supplementary Material). An electrostatic model, based on the interaction lengths calculated from the positions of the small peaks in the force curves like those displayed in Figure 1D (right panels), is proposed in the Supplementary Material.

Notably, these short- and long-distance mechanically weak interactions are observed to be in equilibrium with the "β-like" structures. The population of the former always decreases while that of the latter increases. This result is in accord with the observation by Zhu *et.al.* that a driving force to α-helical structures inhibits αSyn fibril formation[60] and also rule out any template-mediated β-sheet imprinting by the I27 linkers. This conclusion is confirmed by the data on 500 mM Tris/HCl buffered solutions showing that when electrostatic interactions between the αSyn moiety and the flanking I27 linkers are decreased, the population of "β-like" conformers increase. We can also expect entropic effects due to the presence of the flanking I27 domains to drive the protein towards more extended conformations rather than compact conformations [64,65].

These considerations indicate that the design and usage of alternative linkers or experimental strategies may prove useful in the future to further discriminate the effective conformational distribution of αSyn from alterations due to the interaction with the linkers.

**CONCLUSIONS**

For the first time we applied the AFM-based single-molecule mechanical unfolding methodology to a multimodular protein containing the αSyn moiety. This approach brings into play three main methodological capabilities inaccessible to the bulk ensemble-averaged spectroscopies previously applied to study the structure of αSyn and other natively unstructured proteins.

The first is the possibility to work strictly at the single molecule level, thus ensuring that the conformer distribution of the monomeric αSyn is detected and quantified without interference from oligomeric soluble forms of the protein and therefore of any possible intermolecular imprinting towards the amyloidogenic β-structures.

The second capability is that of spanning the conformational space of the protein under investigation and of directly catching and quantifying all of its conformers with a lifetime longer than $10^{-3}$ sec. These conformers, because of their longer life time, might be the most biologically relevant. Three classes of the monomeric αSyn conformations, including random coil, mechanically weakly folded and β-like were characterized by our experiments. They could be detected even in low concentration without the necessity of selectively enhancing one of them by adding specific agents to unbalance the conformational equilibrium, as most commonly done so far with bulk ensemble-averaged experiments.

The third capability is that of following shifts in the population of these classes of conformers in response to changing the solution conditions or the protein sequence and to detect them even if scarcely populated. In the case of αSyn, conditions known to promote oligomerization and aggregation like the presence of $Cu^{2+}$, the A30P mutation, or a radical increase of ionic strength, markedly shift its conformational equilibrium towards the "β-like" form at the expense of other structures. These results indicate that the "β-like" curves contain the signature of the structural precursor to αSyn oligomerization. We suggest that the different aggregation propensities and, ultimately, the pathogenicity displayed by αSyn under different environmental conditions or point mutations can be triggered by unbalance introduced into the delicate equilibria among αSyn conformers.

These capabilities suggest that in the near future single molecule methodologies will play a crucial role in studies of the folding equilibria of the NUPs monomers and in particular, in the detection and quantification of the conformers that can lead to aggregation of those proteins. Our results suggest the feasibility of single-molecule approaches to the testing of novel pharmacological or biophysical therapies for pathologies involving the conformational equilibria of NUPs.

**MATERIALS AND METHODS**

*Polyprotein design and expression.* We followed the protein construct design proposed by J. Fernandez for the study of the random coiled titin N2B segment [41]. Chimeric polyproteins were obtained starting from pAFM1-4, pAFM5-8 and pAFM(I27)3mer vectors, kindly provided by Prof. Jane Clarke (Cambridge University) and constructed according to Ref. [43]. αSyn or its A30P mutated sequences were amplified by PCR using two different pairs of primers, each containing unique restriction sites. A first pair contained KpnI and XbaI sites and a second one SacI and BssHII sites. The original eight I27 module plasmid was reconstituted from pAFM1-4 and pAFM5-8, obtaining pAFM8m vector. pAFM8m was then digested with KpnI and XbaI and ligated to the amplified αSyn sequence, cleaved by the same enzymes in substitution of the two central titin modules to give the pAFM3s3 vector (see Supplementary Material). By a similar strategy, the pAFM(I27)3mer vector was digested with SacI and BssHII, and the central titin module replaced by αSyn sequence, obtaining the pAFM1s1 vector. The obtained expression plasmids, pAFM3s3 and pAFM1s1, code for two chimeric polyproteins composed of a single αSyn module flanked on either side by three tandem I27 domains or by just one, named 3S3 and 1S1, respectively. The two pAFM8m and pAFM(I27)3mer vectors (coding for two recombinant poly(I27) proteins named 8T and 3T) were transformed into *E. coli* C41(DE3) cells [66] (obtained from Professor John E. Walker (MRC – Dunn Human Nutrition Unit, Cambridge) with the agreement of the Medical Research Council center of Cambridge). The cells were grown and the expression of proteins was induced as described in [43]. Recombinant proteins were purified by $Ni^{2+}$-affinity chromatography in 20mM sodium phosphate buffer pH 8, 500 mM NaCl; the elution from the resin was obtained with 20 mM imidazole. After dialysis, proteins were kept at -80°C in PBS with 15% glycerol. The purification gel is shown in Supplementary Material.

*CD Experiments* - CD measurements were carried out on a JASCO J-715 spectropolarimeter interfaced with a PC. The CD spectra were acquired and processed using the J-700 program for Windows®. All experiments were carried out at room temperature using HELLMA quartz cells with Suprasil® windows and an optical path-length of 0.1 cm. Spectra were recorded in the 190-260 nm wavelength range using a bandwidth of 2 nm and a time constant of 2 s at a scan speed of 50 nm/min. The signal-to-noise ratio was improved by accumulating at least four scans. All spectra are reported in terms of mean residue molar ellipticity $[\Theta]_R$ (deg cm$^2$ dmol$^{-1}$).

*Fluorescence experiments* - Fluorescence emission spectra were recorded on a Perkin-Elmer LS 50 spectrofluorimeter equipped with a thermostated cell compartment and interfaced with a personal computer using the FL-WinLab program for Windows®. Sample measurements were carried out using a HELLMA ultra-micro cell with Suprasil® windows and an optical path length of 10 x 2 mm. Fluorescence spectra were obtained at 25 °C using an excitation wavelength of 288, with an excitation bandwidth of 4 nm and emission bandwidth of 4 nm. Emission spectra were recorded between 290-380 nm at a scan rate of 60 nm/min.

*Buffer elemental analysis*. Due to the well known structuring effects of divalent metal ions on αSyn [35], an accurate elemental analysis of the buffer was performed to exclude artifacts in our results due to metal contamination. The high concentration Tris-buffer solution (500 mM) was analyzed for metal contents by atomic absorption spectroscopies. The measured concentrations were Cu = 0.2 ± 0.1 nM, Zn = 3.5 ± 0.1 nM, Fe = 0.9 ± 0.1 nM, and Ca = 22.5 ± 0.1 nM. These values are two orders of magnitude lower than the concentration required to induce structural effects on αSyn [67].

*Surface preparation*. Gold (Alfa Aesar, 99.99%) was deposited onto freshly cleaved mica substrates (Mica New York Corp., clear ruby muscovite) in a high vacuum evaporator (Denton Vacuum Inc., model DV502-A) at 10$^{-5}$ Torr. Before deposition, the mica was preheated to 350 °C by a heating stage mounted behind the mica to enhance the formation of terraced Au(111) domains. The typical evaporation rate was 3 Å/s, and the thickness of the gold films ranged around 300 nm. The mica temperature was maintained at 350 °C for 2 hours after deposition for annealing. This method produced samples with flat Au(111) terraces. These films were fixed to a glass substrate with an epoxy (EPO-TEK 377, Epoxy Tech.). They were then separated at the gold-mica interface by peeling immediately before functionalization with the desired molecules. This procedure produced gold substrates with a flat surface morphology due to the templating effect of the atomically flat mica surface [68,69].

*Force spectroscopy experiments*. For each experiment, a 20 μL drop of 3S3 construct solution (160 μg/mL) was deposited on the freshly peeled gold surface for about 20 minutes. Single molecule force spectroscopy experiments were performed using a commercially available AFM system: Picoforce AFM with Nanoscope IIIa controller (Digital Instruments, Santa Barbara, CA) using V-shaped silicon nitride cantilevers (NP; Digital Instruments, Santa Barbara, CA) with a spring constant calibrated by the thermal noise method [70]. The pulling speed was 2.18 μm/s for all experiments. The buffer used was Tris/HCl (10 mM or 500 mM, pH 7.5; the 10 mM buffer was obtained by diluting the 500 mM buffer with milliQ ultrapure water). For $CuCl_2$ experiments, the protein was deposited in a drop with the addition of a final concentration of 1 μM $CuCl_2$ and left on the surface for about 20 minutes, and the experiments were carried out in 10 mM Tris/HCl with 1 μM $CuCl_2$. Control experiments in DTT were made in 50 mM DTT Tris/HCl buffer.

*Data analysis*. The force curves were analyzed using the commercially available software from Digital Instrument (Nanoscope v6.12r2), custom Origin scripts and *Hooke*, a Python-based home coded force spectroscopy data analysis program (M. Sandal, to be published). Force curves were analyzed fitting each peak with a simple WLC force vs. extension model [45] with two free parameters: the contour length *L* and the persistence length *p* (eq. 1). The I27 modules were characterized in terms of the length of the polypeptide chain extended after each unfolding event.

$$F(x) = \frac{k_b T}{p}\left[\frac{x}{L} + \frac{1}{4\left(1-\frac{x}{L}\right)^2} - \frac{1}{4}\right] \quad \text{(eq. 1)}$$

To assess the statistical validity of the comparison between data obtained in 10 mM Tris/HCl buffer and those obtained in other conditions, standard chi square tests were performed. The differences between the 10 mM Tris data set and the other data sets are significant with $p < 0.01$.

**SUPPORTING INFORMATION**

**Figure S1**. (top panel) Schematic diagram of pAFM3s3 vector obtained from cloning αSyn sequence in J. Clarke's pRSET A modified vector and (bottom panel) representation of the chimeric protein 3S3 coded by the cloned DNA sequence. Titin module numbers refers to the original vector described in the work of Steward et. al . pAFM (I27)3mer and pAFM1s1 vectors and corresponding chimeric protein (respectively 3T, 1S1) can be described with similar diagrams (see article text).

**Figure S2**. Sample of purified 3S3, SDS-10% PAGE and Western blot analysis. **(MW)**, molecular mass markers (Amersham Biosciences); **(lane a)**, final product of immobilized metal affinity chromatography (IMAC) purification of 3S3 chimeric protein; **(lane b)**, Western blotting with anti-His tag Ab; **(lane c)**, Western blotting with anti-αSyn Ab. The arrow indicate a band corresponding to 3S3 protein with an expected molecular weight of 78119 Da which is in good agreement with observed electrophoretical mobility. The lower band that is copurified with 3S3 is recognised only by the anti-His tag Ab, indicating that it is probably an abortive product of translation that does not contain αSyn and, consequently, the last three titin modules with the two final cysteines, necessary for linking the protein to the gold surface. The lower band recognised in lane c is present also in negative controls (data not shown) so it can be considered an aspecific band.

**Figure S3.** Gaussian kernel density estimation of the Mechanically Weak Interaction lengths under different conditions. Briefly, a Gaussian function (kernel) has been centered on each data point for each data set. The sum of the kernels, normalized to have unitary integral, is the KDE plot. Kernel bandwidth $h$ (i.e. the Gaussian kernel standard deviation) was automatically calculated for each data set such that it minimizes the asymptotic mean integrated square error: $h = \sigma \left( \frac{4}{3n} \right)^{\frac{1}{5}}$, where σ is the standard deviation of data and $n$ is the size of the data set. The plot has been calculated using Statistics for Python (http://bonsai.ims.u-tokyo.ac.jp/~mdehoon/software/python/Statistics/).

**Figure S4**. An electrostatic model that may explain mechanically weak interactions. (**A**) Schematic representation of the central portion of 3S3 construct, evidencing the I27 modules flanking the αSyn element (named I27N and I27C, see text). In the αSyn moiety three regions are evidenced: i) the amphipatic region, prone to fold in α-helical structures when in contact with phospholipid membranes; ii) the fibrillogenic NAC region, characteristic of the fibrils core of αSyn amyloid; and iii) the acidic C-terminal tail, strongly charged and not prone to assume any fold. The reported quotes correspond to interactions that may lead to the small peaks observed in curves featuring mechanically weak interactions (see text). The colors of the different regions depict the electrostatic potential: *red* for the negatively charged I27 (see text) and for the acidic C-terminal region of αSyn; *white* for the hydrophobic NAC region of αSyn; and *blue* for the positively charged amphipatic region of αSyn. (**B**) Cartoon (left panels) and solid surface representations of the electrostatic potential (central and right panels) of titin I27 domain obtained using the program DeepView . I27 coordinates were taken from the Protein Data Bank (PDB code: 1WAA). In the cartoon, the secondary structure elements are colored in *blue* for β-strands. Surfaces were colored according to the calculated electrostatic potential contoured from -5.0 $kT/e$ (*intense red*) to +5.0 (where $k$ = Boltzman constant, T = absolute temperature, and $e$ = electron charge) (*intense blue*). The molecular orientation in the central panel is the same as that in the cartoon (left panel), whereas that in the right panel is rotated by 180° about the vertical axis.

**Figure S5**. Mixing αSyn and I27 in solution does not induce αSyn helical folding. CD spectra of: the 3T construct in PBS 1X e glycerol 15% (red), αSyn in PBS 1X e glycerol 15% (black), a mixture of 3T and αSyn (green), a mixture of the two protein without the 3T contribution (blue). αSyn concentration was 10 µM and the ratio with the 3T construct was adjusted in order to have a ratio of one αSyn per two I27 modules. The spectra indicate that to induce the helical conformation in αSyn by I27 a close and constrained contact is necessary. It is very difficult to obtain this very high local concentration condition in a mixing

experiment.

**Protocol S1.** Supplementary results and discussion. Characterization of the rupture events attributed to mechanically weak interaction. An electrostatic model that explains the observed mechanically weak interactions.

**ACKNOWLEGMENTS**

We thank Dr. J. Clarke for kindly providing the vectors for constructs expression, Prof. J.E. Walker for supplying us the *E. coli* C41(DE3) cells. We thank Prof. G. Legname for critical reading of the manuscript as well for advice and support. We thank Dr. Mark R. Cookson, (NIH) for providing the A30P αSyn plasmid.

**Author contributions.** MS and FV equally contributed to this work.

**Funding**. This work was supported by MIUR-FIRB RBNE03PX83/001; MIUR-FIRB Progetto NG-lab (G.U. 29/07/05 n.175); PRIN 2007, EU FP6-STREP program NMP4-CT-2004 - 013775 NUCAN. LB is supported also by PRIN 2005 and FIRB Progetto RBNE03PX83/006.

**Competing interests**. The authors have declared that no competing interests exist.

**Conformational Equilibria of Monomeric α-Synuclein at the Single Molecule Level**

Massimo Sandal, Francesco Valle, Isabella Tessari, Stefano Mammi, Elisabetta Bergantino, Francesco Musiani, Marco Brucale, Luigi Bubacco, and Bruno Samorì

# SUPPLEMENTARY MATERIAL

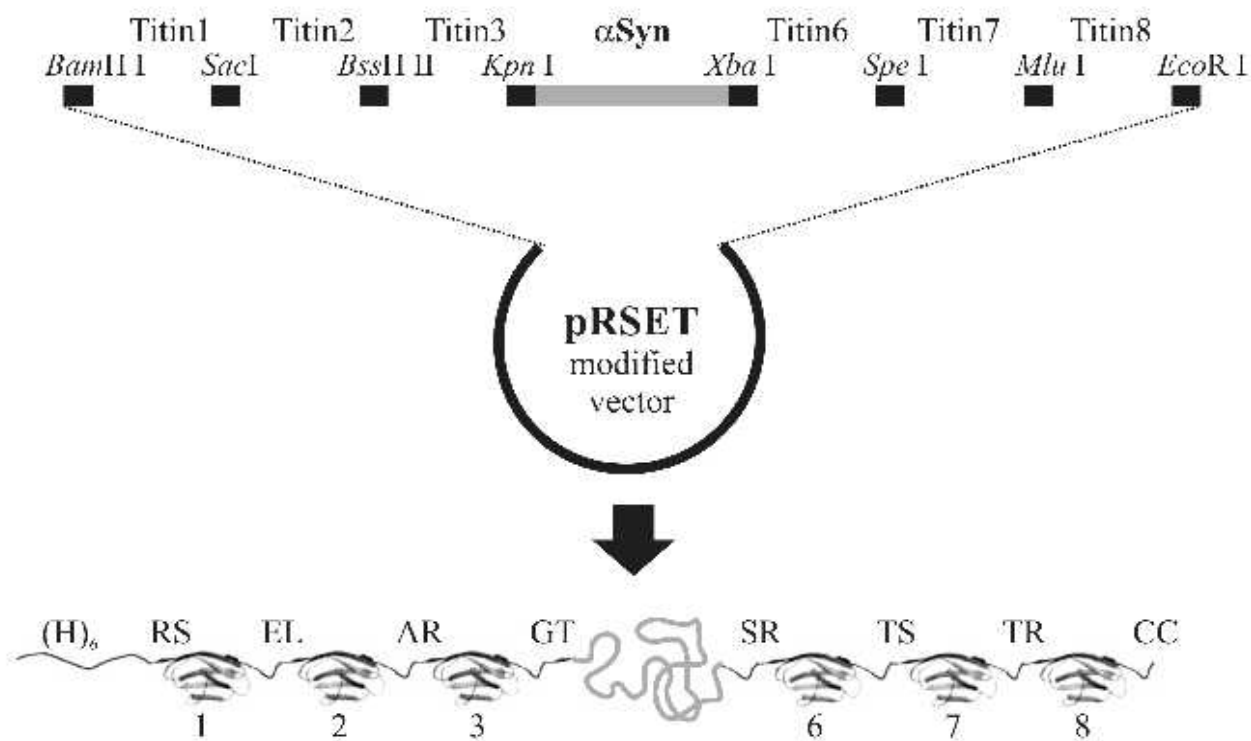

**Figure S1**. (top panel) Schematic diagram of pAFM3s3 vector obtained from cloning αSyn sequence in J. Clarke's pRSET A modified vector and (bottom panel) representation of the chimeric protein 3S3 coded by the cloned DNA sequence. Titin module numbers refers to the original vector described in the work of Steward et. al . pAFM (I27)3mer and pAFM1s1 vectors and corresponding chimeric protein (respectively 3T, 1S1) can be described with similar diagrams (see article text).

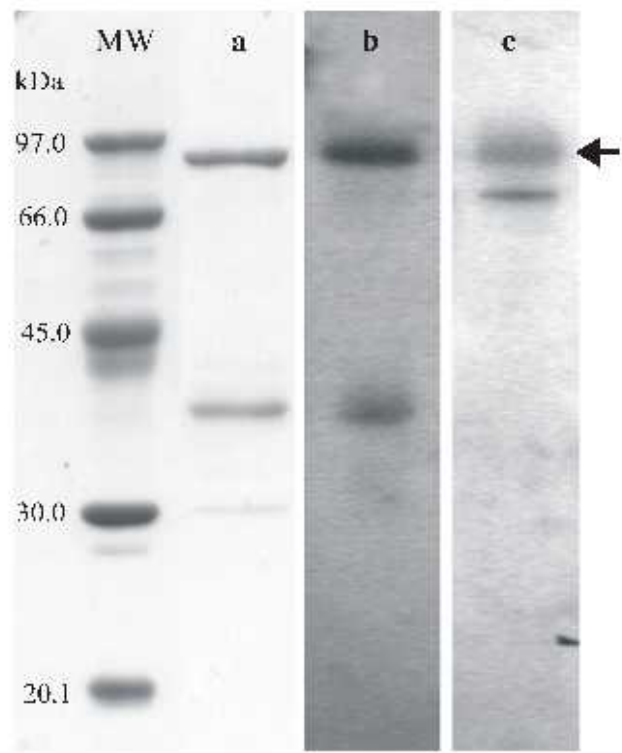

**Figure S2**. Sample of purified 3S3, SDS-10% PAGE and Western blot analysis. **MW**, molecular mass markers (Amersham Biosciences); **lane a**, final product of immobilized metal affinity chromatography (IMAC) purification of 3S3 chimeric protein; **lane b**, Western blotting with anti-His tag Ab; **lane c**, Western blotting with anti-αSyn Ab. The arrow indicate a band corresponding to 3S3 protein with an expected molecular weight of 78119 Da which is in good agreement with observed electrophoretical mobility. The lower band that is copurified with 3S3 is recognised only by the anti-His tag Ab, indicating that it is probably an abortive product of translation that does not contain αSyn and, consequently, the last three titin modules with the two final cysteines, necessary for linking the protein to the gold surface. The lower band recognised in lane c is present also in negative controls (data not shown) so it can be considered an aspecific band.

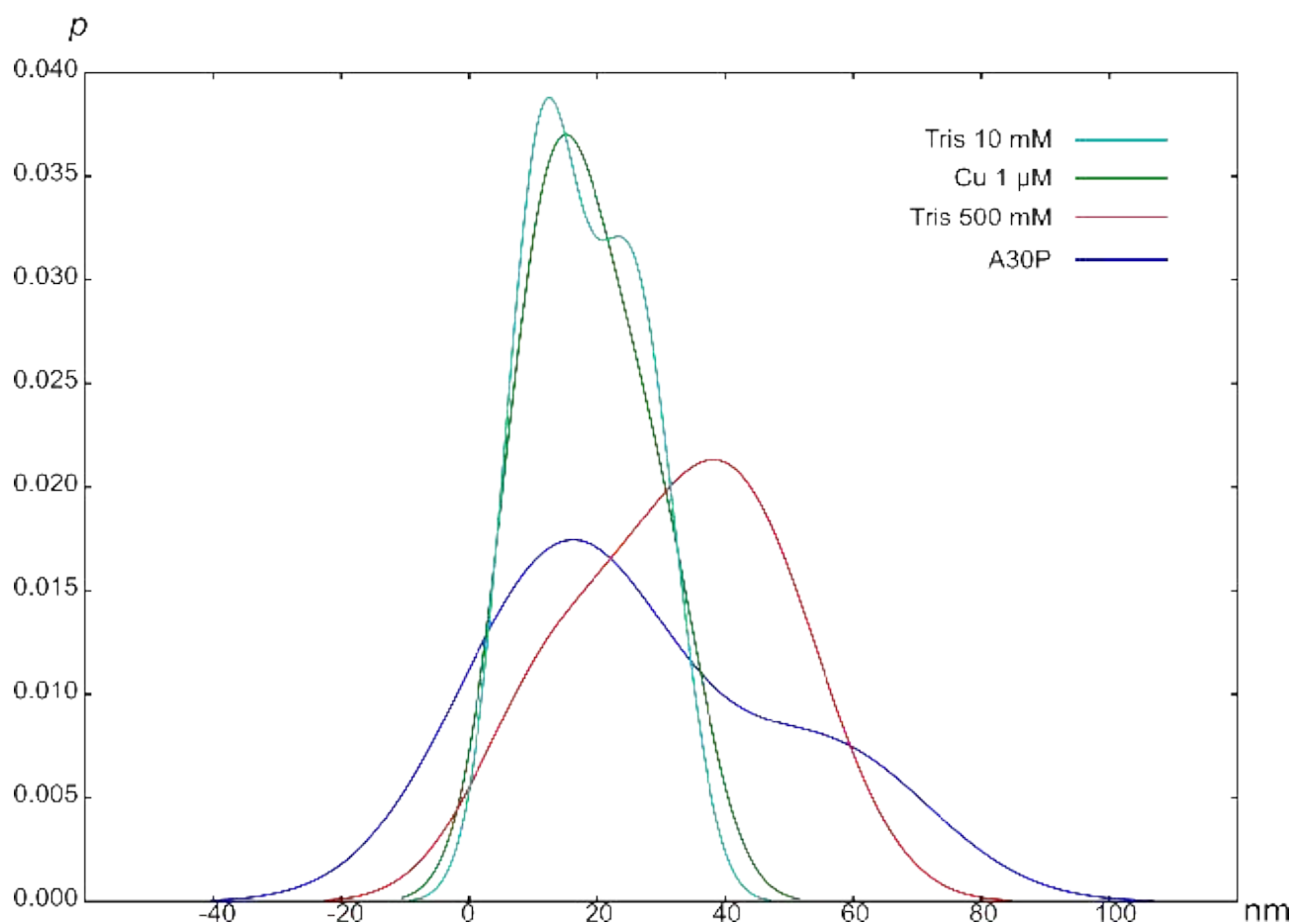

**Figure S3. Gaussian kernel density estimation of the Mechanically Weak Interaction lengths under different conditions.** Briefly, a Gaussian function (kernel) has been centered on each data point for each data set. The sum of the kernels, normalized to have unitary integral, is the KDE plot. Kernel bandwidth *h* (i.e. the Gaussian kernel standard deviation) was automatically calculated for each data set such that it minimizes the asymptotic mean integrated square error: $h = \sigma \left( \frac{4}{3n} \right)^{\frac{1}{5}}$, where σ is the standard deviation of data and *n* is the size of the data set. The plot has been calculated using Statistics for Python (http://bonsai.ims.u-tokyo.ac.jp/~mdehoon/software/python/Statistics/).

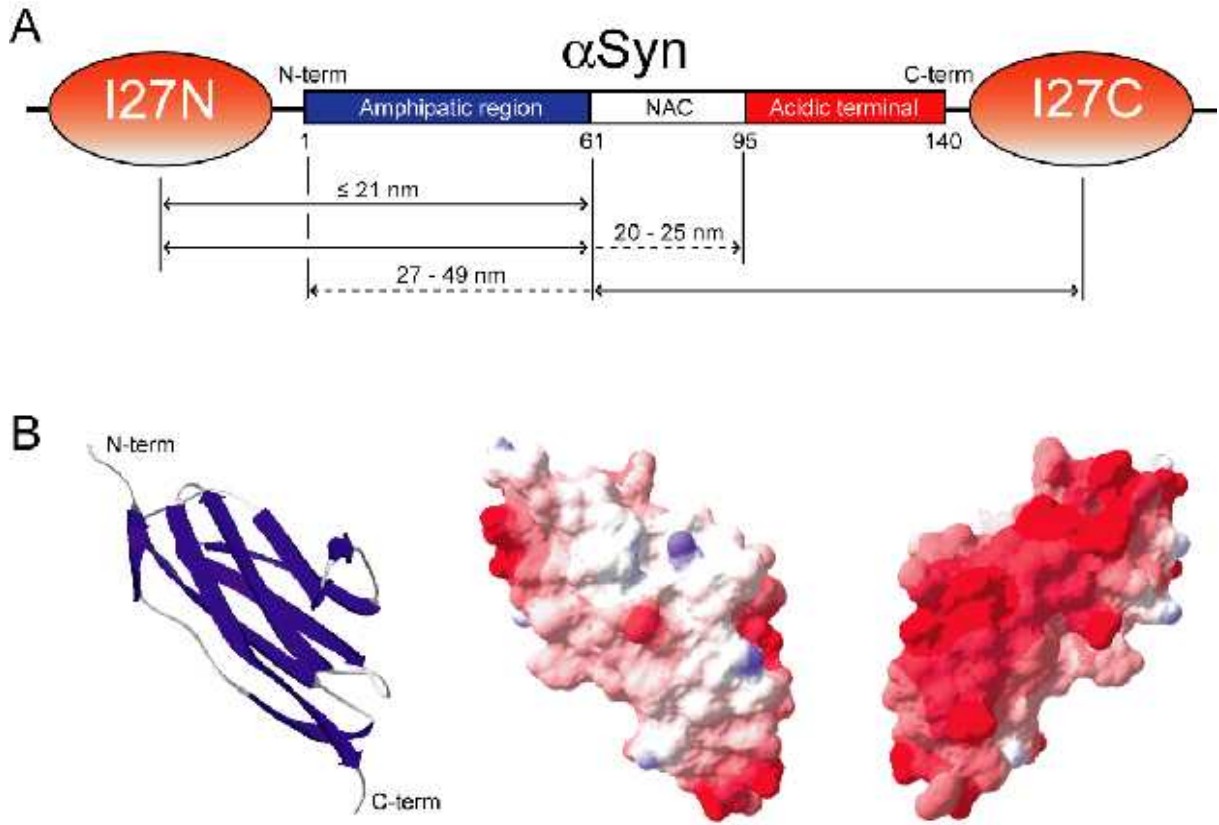

**Figure S4**. **An electrostatic model that may explain mechanically weak interactions.** (**A**) Schematic representation of the central portion of 3S3 construct, evidencing the I27 modules flanking the αSyn element (named I27N and I27C, see text). In the αSyn moiety three regions are evidenced: i) the amphipatic region, prone to fold in α-helical structures when in contact with phospholipid membranes; ii) the fibrillogenic NAC region, characteristic of the fibrils core of αSyn amyloid; and iii) the acidic C-terminal tail, strongly charged and not prone to assume any fold. The reported quotes correspond to interactions that may lead to the small peaks observed in curves featuring mechanically weak interactions (see text). The colors of the different regions depict the electrostatic potential: *red* for the negatively charged I27 (see text) and for the acidic C-terminal region of αSyn; *white* for the hydrophobic NAC region of αSyn; and *blue* for the positively charged amphipatic region of αSyn. (**B**) Cartoon (left panels) and solid surface representations of the electrostatic potential (central and right panels) of titin I27 domain obtained using the program DeepView . I27 coordinates were taken from the Protein Data Bank (PDB code: 1WAA). In the cartoon, the secondary structure elements are colored in *blue* for β-strands. Surfaces were colored according to the calculated electrostatic potential contoured from -5.0 $k$T/$e$ (*intense red*) to +5.0 (where $k$ = Boltzman constant, T = absolute temperature, and $e$ = electron charge) (*intense blue*). The molecular orientation in the central panel is the same as that in the cartoon (left panel), whereas that in the right panel is rotated by 180° about the vertical axis.

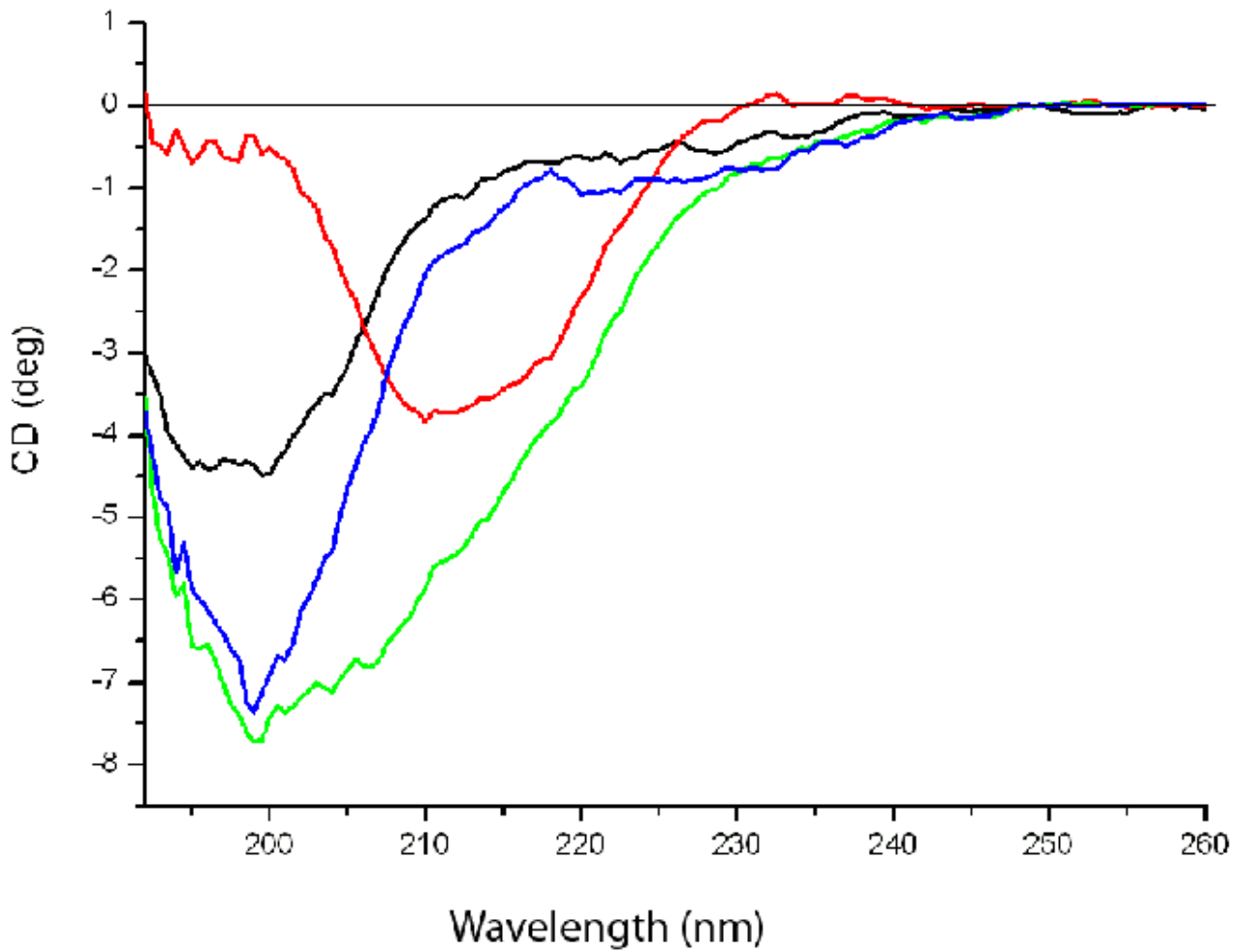

**Figure S5**. **Mixing αSyn and I27 in solution does not induce αSyn helical folding.** CD spectra of: the 3T construct in PBS 1X e glycerol 15% (red), αSyn in PBS 1X e glycerol 15% (black), a mixture of 3T and αSyn (green), a mixture of the two protein without the 3T contribution (blue). αSyn concentration was 10 µM and the ratio with the 3T construct was adjusted in order to have a ratio of one αSyn per two I27 modules. The spectra indicate that to induce the helical conformation in αSyn by I27 a close and constrained contact is necessary. It is very difficult to obtain this very high local concentration condition in a mixing experiment.

**Protocol S1. Supplementary Results and Discussion.**

**The distribution of mechanically weak interaction lengths is differently affected by variations in the environment or protein sequence.**

We characterized the mechanically weak interactions (see text, Fig.1D) by measuring the distance between the contour length of events attributed to them and the contour length measured on the first I27 peak in the same curves, both obtaind by a double-parameter WLC fit. This value should correspond to the chain segment length enclosed by the mechanically weak interactions . The larger set of mechanically weak interactions events is that obtained on Tris 10 mM ($nMWI_{\text{Tris 10mM}}$ = 19). Data recorded in other conditions show a significant decrease in the occurrence of mechanically weak events. The resulting data sets are too small for a significant statistical comparison to be made for each condition, but allow to visualize general trends.

We tried to understand the underlying trends by using kernel density estimation (KDE) to obtain a sketch of the probability density function of interaction lengths. This refined method does not suffer of biases in the choice of binning boundaries and allows for a more objective visualization of small data sets than a classical histogram.
The KDE plot (Supplementary Material Figure 3) of the interaction lengths found in Tris 10 mM and the Tris 10 mM + 1 µM $CuCl_2$ for mechanically weak interactions events are almost perfectly superimposable, despite the difference in the size of the data set ($nMWI_{\text{Cu 1 µM}}$ = 5). The larger number of data in plain Tris 10 mM allows nonetheless to distinguish a bimodal distribution distinguishing mechanically weak interactions enclosing loops of 11 and 25 nm. Experiments conducted with 500 mM buffer concentration ($nMWI_{\text{Tris 500 mM}}$ = 8) and the A30P mutant ($nMWI_{\text{A30P}}$ = 6) show instead a broadening of the KDE and allow to infer a shift from short- to long-distance interactions.

These preliminary data seem to indicate an increase of the conformational freedom of the correspondent architectures. This is in line with the observed increased flexibility of the αSyn chain in presence of the A30P mutation . Instead, as previously discussed, in the case of added $Cu^{2+}$ ions, the reason for the increase of β-like structures lies most probably in the stabilization of the metal binding interface centered around His 50, and not on a broad effect on the protein flexibility. However larger data sets will be required to confirm these conclusions.

**A model of the mechanically weak interactions**

Considering electrostatic and hydrophobic features of the different construct domains, we can try, very tentatively, to associate the mechanically weak interactions with those of different domains within the 3S3 construct. We can sketch and divide the overall structure of αSyn into three main regions: the amphipatic N-terminal tail (aa 1-61), positively charged because of its Lys residues; a hydrophobic central (aa 61-95) region containing the fibrillogenic NAC segment; and the highly negatively charged acidic C-terminal tail. In the 3S3 construct the αSyn N- and C- termini are flanked by partially negatively charged I27 modules (Fig. 5) , that we label I27N and I27C, respectively. Contacts between the negatively charged region of I27N and the positive αSyn N-terminal would appear as peaks that correspond to short-range distances along the primary structure: lower than or equal to 21 nm. The peaks in the middle-range distance of 21-33 nm, and those in the long-range distance of 27-49 nm might correspond to contacts of I27N with the central region of αSyn, and to contacts of the αSyn N-terminal with either the I27C or the αSyn C-terminal, respectively.

The increase of the electrostatic shielding with the buffer concentration has two main effects. First, the population of the mechanically weak interaction featuring structures decreases from about 55% to 23% (see text, Fig. 2), most likely because the interactions with the I27 domains that drive αSyn towards α-helical structures are less favored. Second, the rigidity of the C-terminal tail is decreased as also evidenced by the shift from the short- to the long-distance intermolecular interactions with the increase of the buffer concentration (see Supplementary Material Figure 3). This reduction of rigidity is associated to an increased conformational freedom of the whole αSyn moiety, and most importantly of its NAC region .

The intertwining of both effects can shift the conformational equilibrium towards the β-like structures. In fact, their population increases from about 7% to almost 30% on passing from 10 to 500mM Tris/HCl buffer (Fig. 2A).

It is also possible that mechanically weak interaction events contain the signature of the unlocking of αSyn intramolecular interactions. These latter interactions can sustain the fairly compact structures reported by many authors already , even if their distribution is in our experiments affected by the insertion of the αSyn moiety in the 3S3 construct. For instance the electrostatic repulsion between the negatively charged I27 modules should hinder the long-range contacts between the C- and N-terminal domains proposed by several authors (see below, and Fig. 5) .

**Supplementary References:**